\documentstyle[psfig]{laa}  % LaTeX A&A  Monotype Times Fonts
\begin{document}
   \thesaurus{20         % A&A Section 8: Stars
              (08.01.1;  % Stars: abundances,
               08.03.2;  % Stars: chemically peculiar,
               08.05.1;  % Stars: early type,
                        )
                }
   \title{Variability of the He{\sc i}5876~{\AA} line in early type 
chemically peculiar stars}
\author{G. Catanzaro\inst{1,4}, F. Leone\inst{2,4} \and F. A. 
Catalano\inst{3,4}} 
\offprints{G. Catanzaro - Istituto di Astronomia, Universit\`a di Catania,
Viale A. Doria 6, I--95125 Catania, Italy}
\institute{Dipartimento di Fisica della Materia e Tecnologie Fisiche 
Avanzate, Universit\`a di Messina, Salita Sperone 31, I--98166 Messina, 
Italy;
\\e-mail: gcatanzaro@alpha4.ct.astro.it 
\and
Osservatorio Astrofisico di Catania, Citt\`a Universitaria, 
Viale A. Doria 6, I--95125 Catania, Italy;
\\e-mail: fleone@alpha4.ct.astro.it 
\and
Istituto di Astronomia, Universit\`a di Catania, Citt\`a Universitaria, 
Viale A. Doria 6, I--95125 Catania, Italy;\\ 
e-mail: fcatalano@alpha4.ct.astro.it 
\and 
C.N.R.-G.N.A., UdR di Catania, Citt\`a Universitaria, Viale A. Doria 6, 
I--95125 Catania, Italy}
   \date{Received June 3, 1998; accepted August 3, 1998}
   \maketitle
   \begin{abstract}
%______________________________________ Do not leave a blank line here!
%
%  14.Sep.'90: Demo-Vs.
%_____________________________________ Do not leave a blank line here!
Chemically peculiar stars present spectral and photometric variability with
a single period. In the oblique rotator model, the non homogeneous
distribution of elements on the stellar surface is at the origin of
the observed variations. As to helium weak stars, it has been suggested that
photometric and helium line equivalent width variations are out of phase.
To understand the behaviour of helium in CP stars,  we have obtained time 
resolved spectra of the He{\sc i}5876 {\AA} line for a sample of 16
chemically peculiar stars in the spectral range B3 -- A1 and belonging to
different sub-groups. \\ 
The He{\sc i}5876 {\AA} line is too weak to be measured in the 
spectra of the stars HD\,24155, HD\,41269, and HD\,220825. No variation 
of the equivalent width of the selected He line has been revealed in the 
stars HD\,22920, HD\,24587, HD\,36589, HD\,49606, and HD\,209515. The 
equivalent width variation of the He{\sc i}5876 {\AA} line is in phase 
with the photometric variability for the stars  HD\,43819, HD\,171247 
and HD\,176582. On the contrary it is out of phase for the stars 
HD\,28843, HD\,182255 and HD\,223640. No clear relation has been found
for the stars HD\,26571 and HD\,177003.  
   \end{abstract}
%________________________________________________________________
\section{Introduction}
Among the early type stars of the main sequence various groups of 
chemically peculiar stars (henceforth CP stars) are found. According to 
the {\it General Catalogue of Ap and Am stars} (Renson et al. 1991),
among the 6684 so far known or suspected CP stars, more than half 
(3427) are Am stars (CP1) the remaining 3257 being Bp or Ap. Among 
these only 190 stars belong to the He strong (82) or He weak (108) 
subgroups. Usually helium is underabundant in the coolest CP 
stars and overabundant in the hottest ones. 

CP stars are characterised by spectral and photometric variations with
a common period. In the oblique rotator model, proposed by Stibbs (1950),
chemical elements are not homogeneously distributed on the stellar surface 
and the observed variations are due to the stellar rotation. 
Studying the photometric variability of six helium weak stars,
Catalano \& Leone (1996) found that the equivalent width of the
He{\sc i}4026\,{\AA} varies out of phase with respect to the photometric
variability. 

To investigate the behavior of helium and its relation to the 
spectral and light variability, we have performed time 
resolved spectroscopy of the He{\sc i}5876\AA\,line for a sample 
of 16 CP stars in the B6-A1 spectral range and belonging
to different peculiarity classes.
\begin{table*}[tb]
\caption[]{Average equivalent width $\langle$EW$\rangle$ and relative standard
deviation
$\sigma$ for the observed chemically peculiar stars. Spectral type and 
peculiarity class are from {\it The General Catalogue of Ap and Am 
stars} by Renson et al. (1991). T$_{eff}$ and log g have been computed 
according to Moon \& Dworetsky (1985) and Napiwotzki et al. (1993). N refers 
to the number of spectra, P(d) to the variability period in days with the 
relative reference. 
Source for  V magnitudes and $v_{e} \sin i$ was SIMBAD database.
The {\it ``no line''} remark indicates the
absence of the He{\sc i}5876\,{\AA} line in the spectrum;
{\it ``constant''}, the absence of EW variations; {\it ``in phase''}, the coincidence of
maxima of the EW and photometric variations; and {\it ``out of phase''}, the coincidence of
maxima of the EW variation with the minima of photometric variations.}
\label{listcp}
\begin{tabular}{rclcrrcccrlll}
\hline
HD & HR & Sp. Type & V & $v_{e} \sin i$ & T$_{\rm eff}$ & log g & N &
 $\langle$EW$\rangle$ & $\sigma$ & Remarks & ~~~~P(d) & Reference \\
\hline
\hline
22920 & 1121 & B8 Si   & 5.53 & 120~~ & 13700 & 3.72 &  8 & 125 & 13 & constant &  & \\ 
24155 & 1194 & B9 Si   & 6.30 &  50~~ & 13700 & 3.96 &  9 &     &    & no line &  &  \\ 
24587 & 1213 & B6      & 4.65 &  40~~ & 14100 & 4.23 & 18 & 325 & 20 & constant & & \\
26571 & 1297 & B8 Si   & 6.12 &  29~~ & 13000 & 3.16 & 18 & 145 & 15 &         & 1.0646 & Winzer (1974) \\
28843 & 1441 & B9 He wk& 5.81 &  70$^*$~ & 15300 & 4.17 & 21 & 70  & 20 & out of phase & 1.373813 & Mathys et al. (1986) \\
36589 & 1860 & B7      & 6.18 &  90~~ & 14000 & 3.97 & 9  & 350 & 8  & constant &  &  \\
41269 & 2139 & B9 Si   & 6.20 &  75~~ & 10800 & 3.82 & 10 &     &    & no line &  &  \\
43819 & 2258 & B9 Si   & 6.30 &  14~~ & 11100 & 3.66 & 12 & 95  & 30 & in phase & 15.0305 & Adelman (1997) \\
49606 & 2519 & B8 HgMnSi& 5.85&  35~~ & 13100 & 3.83 & 14 & 145 &  7 & constant &  &  \\ 
171247& 6967 & B8 Si   & 6.42 &  60~~ & 11300 & 3.40 & 23 & 95  & 15 & in phase & 3.9124 & North (1992) \\
176582& 7185 & B5 He wk & 6.41 & 145~~ & 18300 & 4.31 & 12 & 360 & 30 & in phase & 1.58175 & This work \\
177003& 7210 & B3 He   & 5.37 & 40~~ & 18700 & 4.11 & 14 & 705 & 25 & & 1.835 & This work \\
182255& 7358 & B6 He wk & 5.13 & 45~~ & 14400 & 4.17 & 18 & 325 & 15 & out of phase  & 1.26263 & This work \\
209515& 8407 & A0 CrSiMg & 5.60 & 100~~ & 9600 & 3.73 & 7 & 40 & 5 & constant &  & \\
220825& 8911 & A1 CrSrEu & 4.94 & 30~~ & 10400 & 4.46 & 7 &   &   & no line &  & \\
223640& 9031 & B9 SiSrCr & 5.19 & 20~~ & 12400 & 3.62 & 6 & 60 & 15 &  out of phase & 3.735239 & North et al. (1992)\\
\hline
\multicolumn{13}{l}{\rm $^*$The projected rotational velocity of HD\,28843 has been here measured.}\\
\end{tabular}
\end{table*}
\section{Observations and data analysis}
For the chemically peculiar stars listed in 
Table~\ref{listcp} echelle spectra were obtained in 1995 at the 2.1~m 
telescope of the Complejo Astron\'omico El Leoncito equipped with a 
Boller \& Chivens cassegrain spectrograph and in 1997 at the 
91~cm telescope of the Catania Astrophysical Observatory equipped with 
a Czerney-Turner echelle spectrograph.  \\
The data were analysed by using IRAF package. The lines of the 
wavelength calibration lamp show that R=16000 for the 1997 data set and 
R=13000 for the 1995 data set. The achieved S/N was between 100 and 200.
When possible, equivalent widths were measured by a Gaussian fit 
of spectral lines after having removed possible continuum slope; otherwise 
a measure of the area between the line profile and the 
continuum was obtained. Following Leone et al. (1995), we 
estimated the error in the measured equivalent width with the relation: 
\begin{equation}
\Delta W = \frac{1}{2} \left(2 \frac{v_e \sin i}{c} \lambda\right)
\frac{1}{S/N}
\label{errorbar}
\end{equation}
where the quantity in brackets is the total extension of the line as 
deduced from the rotational broadening. Adopted $v_e \sin i$ values (Table\,1)
are from SIMBAD with the exception of HD\,28843 whose projected rotational
velocity was measured from the unblended Si{\sc ii}5865\,{\AA} line. \\
The initial ephemeris of program stars were taken from Catalano \& 
Renson (1984, 1988, 1997), and Catalano et al. (1991, 1993), and 
references therein. If necessary, periods were established using
our spectral observations and Hipparcos photometry\footnote{
The Hipparcos filter, referred to as Hp, extends from 3550 
{\AA} to 8900 {\AA} with the maximum at 4350 {\AA}. The typical 
accuracy of Hipparcos measurements, at the 8th magnitude, is given as 
0.0015 mag (ESA, 1997).}.
A least squares fit of measured EW's and H$_{\rm p}$ magnitudes has
been performed by adopting the function:
\begin{eqnarray}
 &  & A_0 + A_1 \sin (2 \pi (t-t_0)/P + \phi_1) \nonumber \\
 &  & \qquad \qquad \qquad \qquad + A_2 \sin (4 \pi (t-t_0)/P + \phi_2) 
\end{eqnarray}
where $t$ is the JD date, $t_0$ 
is the assumed initial epoch, $P$ is the period in days. A sine wave and 
its first harmonic appear to be quite adequate functions to describe the light 
curves and the spectral variations (North 1984, Mathys \& Manfroid 1985).   
The error in the period value has been evaluated according to the relation 
given in Horne \& Baliunas (1986).

As to the coolest CP stars, the effective temperatures and gravities have
been determined by means of {\it ad hoc} Napiwotzki et al. (1993) relations.
As to helium peculiar stars, Hauck \& North (1993) found that {\it classical}
methods are still reliable to determine their effective temperature. Thus,
we have used the Moon \& Dworetsky (1985) grids as coded by Moon (1985). 
The source of Str\"omgren photometry was SIMBAD.

To ascertain if the selected stars present a peculiar helium abundance
we have compared the measured equivalent widths of the He{\sc i}5876{\AA} line
with the NLTE computations of Leone \& Lanzafame (1998) for solar
composition stars with $\log$\,g = 3.5, 4.0 and 4.5 and 9000 K 
$<$ T$_{\rm eff} <$ 19000 K.

\section{Individual stars}  
\subsection{HD\,22920 (= HR\,1121 = 22 Eri)}
According to Maitzen (1976), the silicon star HD\,22920 has a low value
of the photometric peculiarity index $\Delta a$ (= 0.011). Photometric 
observations have been carried out by Bartholdy (1988) who found this star 
to be variable with a period of 3.95 d. North (1990 priv. comm.) 
found two possible periods almost equally probable: 3.96 d, very close 
to Bartholdy's (1988), and 1.33 d.

No evidence of variability has been found in our spectra of the
He{\sc i}5876{\AA} line. The mean value 
of the equivalent width is: $\langle$EW$\rangle$ = 
125 $\pm$ 13 m{\AA}. The effective temperature of HD\,22920 resulting
from Napiwotzki et al. (1993) relation is T$_{\rm eff}$ = 13700 K.
Figure\,9 shows that the He{\sc i}5876{\AA} line equivalent width of HD\,22920
is smaller than expected for a main sequence star of the same effective 
temperature.   

\subsection{HD\,24155 (= HR\,1194 = V766 Tau)}
The UBV photometric variability of HD\,24155 has been studied by Winzer 
(1974), who reported a possible period of 2.5352 d. Renson \& Manfroid (1981)
found P = 2.53465~$\pm$~0.00015~d. The observed light 
curves show a quite large amplitude (0.10 mag) with very sharp minima 
and quite broad maxima, hence this star is the fourth largest amplitude 
silicon star known, exceeded only by HD\,215441, CU Vir and HR\,7058. 

Assuming Renson \& Manfroid's (1981) period, our nine spectra are well
distributed in phase, but none of them shows a measurable He{\sc i}5876
{\AA} line. Because of its  effective temperature of 13700 K, HD\,24155 is 
an extremely helium weak star (Fig.\,9).

\subsection{HD\,24587 (= HR\,1213 = $\tau^8$ Eri)}
HD\,24587 is listed in the {\it General Catalogue of Ap and Am stars} 
by Renson et al. (1991) as a suspected CP star. 
Feinstein (1978) used this star as standard for his measurements of 
hydrogen lines in He weak stars. HD\,24587 has in fact been considered 
as a standard for {\it uvby} (Garnier 1972 - personal communication to 
Mathys et al. 1986) and {\it $\beta$} photometry (Strauss \& Ducati 1981). 
Mathys et al. (1986) found this star to be a light variable with a period 
of 1.728 d and concluded that the light curves resemble those of many CP 
stars. Recently Leone \& Catanzaro (1998) have performed a 
spectroscopic study and concluded that this star presents chemical 
elements which are slightly underabundant with respect to main sequence 
stars.

Our measurements of the He{\sc i}5876 {\AA} line do not show any
variation of the equivalent width; the mean value is: $\langle$EW$\rangle$ = 
325 $\pm$ 20 m{\AA}. From Moon's algorithm we find that T$_{\rm eff}$ = 
14100 K and the EW of the He{\sc i}5876 {\AA} line is close to the value 
expected for a main sequence star (Fig.\,9). 
These facts confirm Leone \& Catanzaro's (1998)  conclusion that 
HD\,24587 is not a peculiar star. 

\subsection{HD\,26571 (= HR\,1297)}
The peculiarity of HD\,26571 was first noted by Gulliver (1971) and 
independently confirmed by Bond (1972). On the basis of his spectra, 
Gulliver (1971) described this star as a spectrum variable.
Photometric observations were obtained by Winzer (1974), 
who found HD\,26571 to vary with a period of 1.0646 d.

\begin{figure}[ht]
  \psfig{file=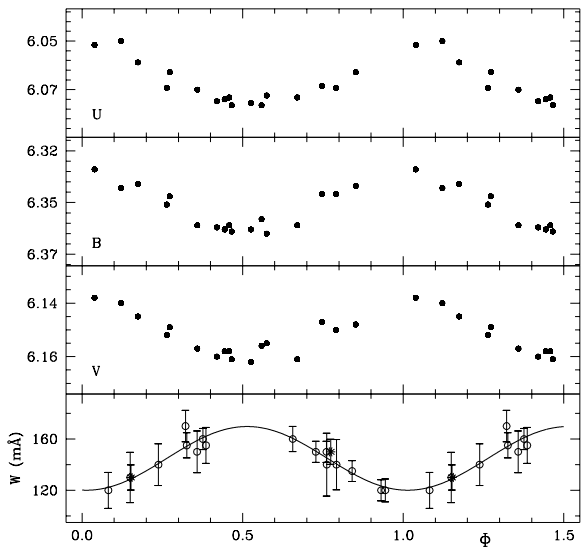,width=8cm}
 \caption{Equivalent width variations of HD\,26571. The errorbar is 
equal to the error in the equivalent width as given by Eq.\,(1). 
Photometric data are from Winzer (1974). Phases have been computed by Eq.\,(3).
Open circles refer to the 1997 data, stars to the 1995 data.
The fitting of EW's performed by Eq.\,(2) is reported with a solid line.
The phase relation between EW and photometric variations may be not real
because of the large time gap between both data sets and the relatively
low accuracy of the variability period.}
  \label{hd26571}
 \end{figure}

Figure~\ref{hd26571} shows the EW variation of the He{\sc i}5876 {\AA} line 
versus the phase computed assuming the initial epoch coincident with the  
light maximum as given by Winzer (1974): 
\begin{equation}
 JD(UBV \; max.) = 2441246.81 + 1.0646 E .
\end{equation}

Winzer (1974) has not published the uncertainty in the period determination, 
hence we have estimated the error by applying the Horne \& Baliunas (1986) 
relation to our data and have found that equivalent widths are phased with 
an expected error $\Delta \Phi$ = 0.3. This means that no phase relation 
can be determined between the photometric and our spectral variations.  

\subsection{HD\,28843 (= HR\,1441 = DZ Eri)}
\begin{figure}[ht]
  \psfig{file=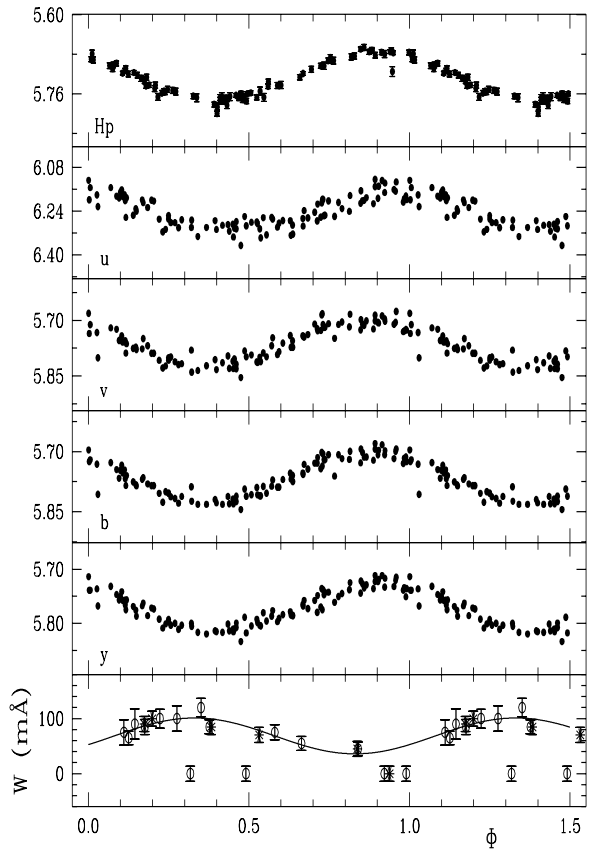,width=8cm,height=12cm}
  \caption{Equivalent width variations of HD\,28843. The bar is 
equal to the error in the equivalent width as given by Eq.\,(1).
Open circles refer to the 1997 data, stars to the 1995 data.  
Photometric observations are from Hipparcos (ESA, 1997) and from 
the ESO LTPV Project (Manfroid et al. 1994, Sterken et al. 1995).}
  \label{hd28843}
 \end{figure}

HD~28843 was classified as B9IV Si He-wk by Davis (1977) and it is classified
as B9 He wk in the {\it General catalogue of Ap and Am Stars}. 
The photometric variability of HD\,28843 had been detected for the first 
time by Cousins \& Stoy (1966) while its peculiar character had been 
confirmed by Jaschek et al. (1969). Photometric observations of this 
star have been carried out by Pedersen \& Thomsen (1977) who found 
variability with a period of 1.374 $\pm$ 0.006 d. This value was improved 
by Pedersen (1979) to the value 1.37375 $\pm$ 0.00035 d.
Manfroid et al. (1984) also used Pedersen \& Thomsen's (1977) data to improve 
the period, their most probable value being 1.373813 $\pm$ 0.000012 d. 
Mathys et al. (1986) concluded that the ambiguity in the choice of the 
best peak in the periodogram could be removed by inclusion of the 
measurements of Dean (1980), confirming the value obtained by Manfroid 
et al. (1984). Further photometric observations have been carried out 
by Waelkens (1985), by the team of the ESO Long-Term Photometry of 
Variable Project (Manfroid et al. 1994, Sterken et al. 1995), and 
by the team of Hipparcos (ESA, 1997). \\
Our spectroscopic data are plotted in Fig.~\ref{hd28843}, versus the 
phase computed from the ephemeris elements of Mathys et al. (1986): 
\begin{equation}
JD(uvby \; max) = 2442777.5 + 1.373813 E
\label{ephe}
\end{equation}
The amplitude of the equivalent line width variations is of the order 
of 75 m{\AA}. From Fig.~\ref{hd28843} a clear anti-correlation is evident 
between the He{\sc i}5876 {\AA} equivalent line width and all the 
Hipparcos and $uvby$ light curves, in the sense that light minima occur 
at the phase of maximum He{\sc i}. Because of the period error determined by 
Manfroid et al. (1984), the expected phase error in our EW variations is 
$\Delta\Phi$ = 0.03. EW variations of the He{\sc i}5876 {\AA} line are then 
out of phase with respect to light variations. \\
Even if most of our equivalent widths periodically vary with the 
ephemeris computed with Eq.~(\ref{ephe}), we  have found several 
(5 out of 21) spectra where the He{\sc i}5876 line is absent (Fig.\,2). 

\subsection{HD\,36589 (= HR\,1860)}
In the {\it General Catalogue of Ap and Am stars}, the star HD\,36589 is a 
suspected CP star.  
Bossi \& Guerrero (1989) and Hao et al. (1996) have used it as 
a comparison star for photometric observations. 
Leone \& Catanzaro (1998) derived chemical abundances and found that
HD\,36589 shows nearly solar values and no evidence of spectral
variability. 

From our spectra we confirm this result: no evidence of variation is 
found in the He{\sc i}5876 {\AA} equivalent line width. On the hypothesis that
HD\,36589 is not a CP star, we have determined T$_{\rm eff}$ = 14000 K
by mean of Moon's relations and found that the average value of
the equivalent widths ($\langle W \rangle$ = 350 $\pm$ 8 m{\AA}) is very
close to that of normal main sequence stars of the same spectral type
(Fig.\,9). 

\subsection{HD\,41269 (= HR\,2139)}
This star has been classified as B9p by Cowley et al. (1969) who described 
it as a mild silicon star. On the basis of a single observing run in UBV, 
Winzer (1974) found a period of 1.68~d, although he could not rule out 
the resonance period of 2.47~d, because of the few observed points.\\ 
In our spectra the He{\sc i}5876 {\AA} line is too weak to 
be measured. According to Napiwotzki et al. (1993) relation,
T$_{\rm eff}$ = 10800 K. Figure~\ref{summary} shows that the helium 
abundance is lower than the expected value for a main sequence star 
of this temperature.

\subsection{HD\,43819 (= HR\,2258 = HIP\,30019 = V 1155 Ori)}
Cowley (1972) classified this star as B9IIIp Si. Photometric measurements 
of HD\,43819 were performed in the UBV system by Winzer (1974) who found a 
light variation with a period of 1.0785~d. Later on Maitzen (1980) found 
the light variation to occur with two possible periods: 0.93~d and 
1.077~d. A spectroscopic study of this sharp lined star ($v_{e} \sin i = 
14$~km~s$^{-1}$, Wolff \& Preston 1978) was carried out by Lopez-Garcia 
\& Adelman (1994), who found iron peak elements ten times overabundant 
and rare earths 1000 times overabundant with respect to solar values. 
From photometric $uvby$ observations Adelman (1997) has deduced a 
period of 15.0305$\pm$0.0003~d, longer than Winzer's (1974) and more 
consistent with the low rotational velocity of this star. This period is 
also confirmed by the Hipparcos observations (Fig.\,3).

\begin{figure}[t]
  \psfig{file=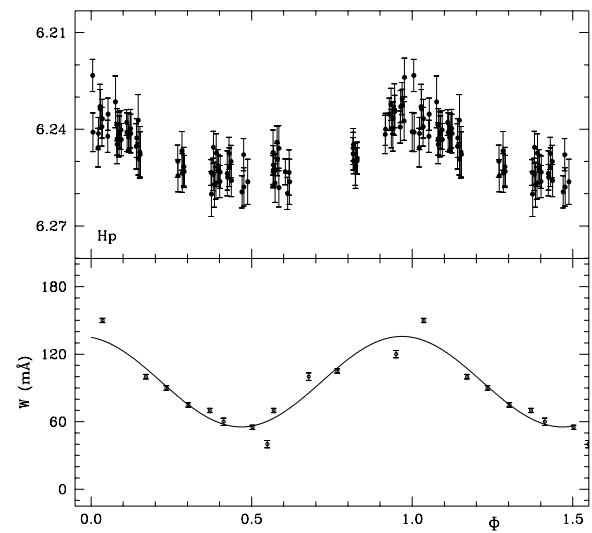,width=8cm}
  \caption{Equivalent width variations of HD\,43819. Errorbar is equal 
to the error in the equivalent width as given by Eq.~(\ref{errorbar}). 
Photometric data are from Hipparcos (ESA, 1997).}
  \label{hd43819}
 \end{figure}

Our He{\sc i}5876 {\AA} equivalent line widths are plotted 
in Fig.~\ref{hd43819} versus the phase computed by means of Adelman's 
(1997) ephemeris elements: 
\begin{equation}
JD(U \; max) = 2441254.16 + 15.0305 E
\end{equation}
The observed EW variation has an amplitude of the order of 80~m{\AA}. 
From Fig.~\ref{hd43819} we see a clear in-phase correlation between light 
and spectral variations. This correlation is expected to be real, 
the phase error being $\Delta\Phi\sim$ 0.01. 

\subsection{HD\,49606(= HR\,2519 = 33 Gem)}
The star HD\,49606 is classified as a B8HgMnSi star by Renson et al. (1991).
Photometric observations of HD\,49606 were performed by Chunakova et al. 
(1981), who found the light variations to occur with a period of 3.099~d, 
and by Glagolevskii et al. (1985), who found two possible period values, 
namely 3.3546~d and 1.41864~d.

The He{\sc i}5876~{\AA} equivalent line width observed in our spectra 
does not show any detectable variation, so that we consider this line 
does not vary with time. The average equivalent width is $\langle W 
\rangle = 145 \pm 7$~m{\AA}. This result confirms the one obtained by
Hubrig \& Launhardt (1993) who searched for variations in the
equivalent width of helium and some metallic lines and did not find any 
evidence of variability.\\
According to an elemental abundances analysis performed by Adelman \& al. 
(1996) our observations show that helium is 
underabundant with respect to solar composition (Fig.\,\ref{summary}).

\subsection{HD\,171247 (= HR\,6967 = HIP\,90971)}
The photometric variability of this star was detected by North 
(1992) who found the period to be 3.9124~d. This value of 
the period is confirmed by the Hipparcos photometry (1997) 
which gives an error on the period equal to 0.0004 d applying
Horne \& Baliunas (1986) formula. 

\begin{figure}[t]
  \psfig{file=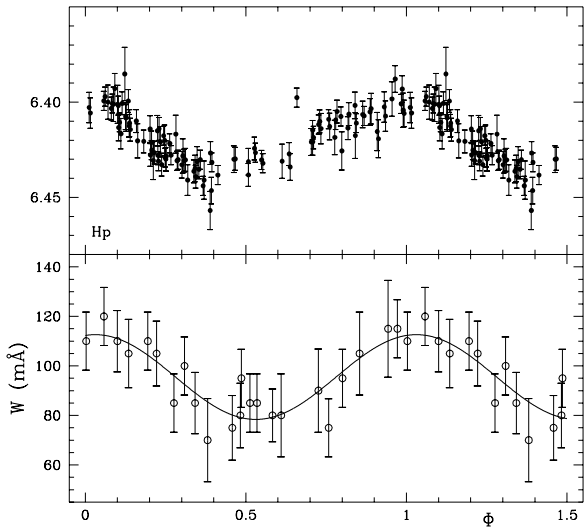,width=8cm}
  \caption{Equivalent width variations of HD\,171247. the bar is equal 
to the error in the equivalent width as given by Eq.~(\ref{errorbar}). 
Photometry is from Hipparcos (ESA, 1997).}
  \label{hd171247}
 \end{figure}

Computing the phase of the measured equivalent widths by means of
North's (1992) ephemeris: 
\begin{equation}
JD([U]{\rm \, Geneva\, max}) = 2447178.245 + 3.9124 E 
\end{equation}
we find a sinusoidal variation of the He{\sc i}5876{\AA} line strength 
(Fig.\,\ref{hd171247}) with an amplitude of the order of 45~m{\AA}. \\
Converting the period error to a phase error, we get $\Delta\Phi$ = 0.07.
We can thus conclude that H$_{\rm p}$ and EW variability are in phase for
the silicon star HD\,171247.  

\subsection{HD\,176582 (= HR\,7185 = HIP\,93210)}
This star is classified as a silicon star (Renson et al. 1991).
Spectroscopic observations of the He{\sc i}4026~{\AA} line strength were
carried out by Pedersen (1976), who found a variation 
with the period 0.8143~d. The period is not representative of
the variability of Hipparcos photometry and He{\sc i}5876~{\AA} equivalent
width.

\begin{figure}[t]
  \psfig{file=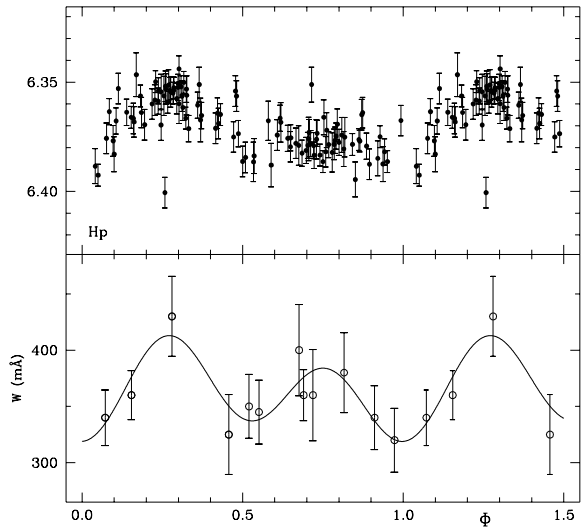,width=8cm}
  \caption{Equivalent width variations of HD\,176582. Errorbar is equal 
to the error in the equivalent width as given by Eq.~(\ref{errorbar}). 
Photometry is from Hipparcos (ESA, 1997).}
  \label{hd176582}
 \end{figure}

By using our spectroscopic data and Hipparcos photometry we found a
period of 1.5817$\pm$0.0003~d. The observations are 
plotted in Fig.~\ref{hd176582} versus the phase computed by means of 
the ephemeris elements: 
\begin{equation}
JD(EW \; min.) = 2450624.6410 + 1.5817 E
\end{equation}
From this figure we see that both curves show a clear evidence of a 
double-wave variation. The observed EW amplitude is of the order 
of 40~m{\AA}.  
The expected phase error is $\Delta\Phi$ = 0.05, and the Hipparcos 
photometry appears to vary in phase with the equivalent width variations
of the He{\sc i}5876~{\AA} line.

\subsection{HD\,177003 (= HR\,7210 = HIP\,93299)}
Sch\"oneich \& Zelwanowa (1984) from their photometric observations
in the UV filters found two possible periods: 0.66~d and 2.1~d.
From UBVRI photometric observations, Vet\"o (1993) found this star to 
be light variable with a period of 0.724~d and amplitudes of about 0.1 
mag. in all filters. An analysis of Hipparcos photometric data does 
not give a clear variability period. 

\begin{figure}[t]
  \psfig{file=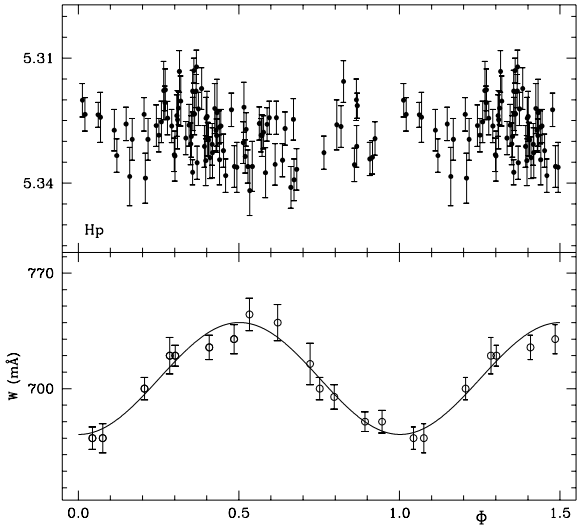,width=8cm}
  \caption{Equivalent width variations of HD\,177003. Errorbar
is equal to the error in the equivalent width as given 
by Eq.~(\ref{errorbar}).}
  \label{hd177003}
 \end{figure}

Our spectroscopic data are not consistent with the periods given
in the literature. A period search of our data yelds two possible 
values: 1.835$\pm$0.004~d and 2.186$\pm$0.005~d. The He{\sc i}5876\,{\AA} line 
equivalent width variation is plotted in Fig.~\ref{hd177003} versus the phase 
computed by means of the ephemeris elements: 
\begin{equation}
JD(EW \; max.) = 2450629.4099 + 1.835 E
\end{equation}
where we have adopted the shorter value of the period which has a smaller 
$\chi^{2}$ value. The variation shown in Fig.~\ref{hd177003} has an
amplitude of the order of 75~m{\AA}.
The photometric variability is not clear for the H$_{\rm p}$ filter 
assuming this period (Fig.\,6) and no conclusion can be drawn concerning a 
possible phase relation between photometric and spectral variations 
of the He{\sc i}5876\,{\AA} line.

\subsection{HD\,182255 (= HR\,7358 = HIP\,95260 = 3 Vul)}
\begin{figure}[t]
  \psfig{file=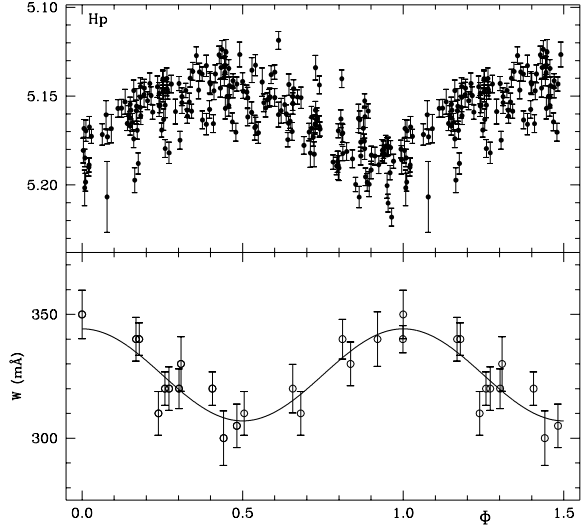,width=8cm}
  \caption{The light and He{\sc i}5876~{\AA} equivalent line width 
variations of HD\,182255. Errorbar is equal to the error in the 
equivalent width as given by Eq.~(\ref{errorbar}). Photometry is from 
Hipparcos (ESA, 1997)}
  \label{hd182255}
 \end{figure}

According to Hube \& Aikman (1991) this star is a nonradial pulsator. It 
has also been observed by Hipparcos, from whose photometry a period of 
1.26239~d has been derived. However this value of the period is not 
perfectly consistent with our spectroscopic observations; instead, 
by using both sets of data the most probable value appears to be 
1.26263 $\pm$ 0.00005~d. Adopting this period, the measured EW of
the He{\sc i}5876\,{\AA} line are plotted in Fig.~\ref{hd182255}
versus the phase computed by means of the ephemeris elements: 
\begin{equation}
JD(EW \; max.) = 2450650.4729 + 1.26263 E 
\end{equation}
The observed EW amplitude is of the order of 65~m{\AA}. 
Since the period error corresponds to a phase error $\Delta\Phi$ = 0.09,
the reported out of phase relation between photometric and helium line
variations is expected to be real (Fig.\,7).  

\subsection{HD\,209515 (= HR\,8407 = V1942 Cyg)}
This star was classified as A0p by Osawa (1965) and as A0 IV by Cowley 
et al. (1969). 
From his photometric observations, Winzer (1974) found a period of 
0.63703~d, concluded that the observed photometric variation is 
typical for a silicon star and suggested that the correct classification 
should be A0p Si. \\
The equivalent width of the He{\sc i}5876\,{\AA} line of the cool CP
star HD\,209515 is constant: 40$\pm$5~m{\AA}. This value of equivalent
width is consistent with the helium abundance of a main sequence star
(Fig.\,9).

\subsection{HD\,220825 (= HR\,8911 = $\kappa$ Psc)}
The variability of HD\,220825 had been detected for the first time by 
Rakosch (1962) who found a period of 0.5805 d. Recently, Ryabchikova et 
al. (1996) determined the period to be 1.418 d and magnetic observations 
performed by Borra \& Landstreet (1980) are also consistent with this 
value.

The He{\sc i}$\lambda$5876~{\AA} line is too weak to be measured.
Assuming T$_{\rm eff}$ = 10400 K, Fig.\,9
shows that helium is underabundant in HD\,220825 with respect to main
sequence stars. 

\subsection{HD\,223640 (= HR\,9031 = HIP\, 117629 = 108 Aqr = ET Aqr)}
The photometric variability of this star has been studied by several 
authors. Morrison \& Wolff (1971) found HD\,223640 to be variable in 
the Str\"omgren system with a period of 3.73 d and noted that light 
curves show quite the same behaviour in all filters. Spectroscopic 
observations were carried out by Megessier \& Garnier (1972) who found 
strongly variable the Ti and Sr lines and constant the Fe lines. 
Moreover the Ti lines correlate with photometric variations in the 
sense that Ti lines are strongest when the star is brightest. This 
correlation has been interpreted by Megessier (1974, 1975) in terms of the 
oblique rotator model taking also into account the sign changes of the 
magnetic field measurements by Babcock (1958).  
Photometric observations in the Geneva system have been performed by
North et al. (1992), they found a period of 3.735239 $\pm$ 0.000024 d which
is consistent with the magnetic data. This period has been confirmed by 
photometric observations in the uvby system performed by Adelman \& Knox 
(1994) and Adelman (1997). \\
According to North et al. (1992), we phased the measured equivalent widths
of the He{\sc i}5876{\AA} line by means of the ephemeris elements:
\begin{equation}
JD(uvby \; max.) = 2444696.820 + 3.735239 E
\end{equation}
The period uncertainty corresponds to a phase error $\Delta\Phi$ = 0.004.
There is evidence of an anti-correlation between light and spectroscopic 
curves. The He{\sc i} line is strongest in coincidence with the light minimum:  
the helium distribution on the surface of HD\,223640 is then not coincident 
with the Ti distribution.  
 \begin{figure}[t]
  \psfig{file=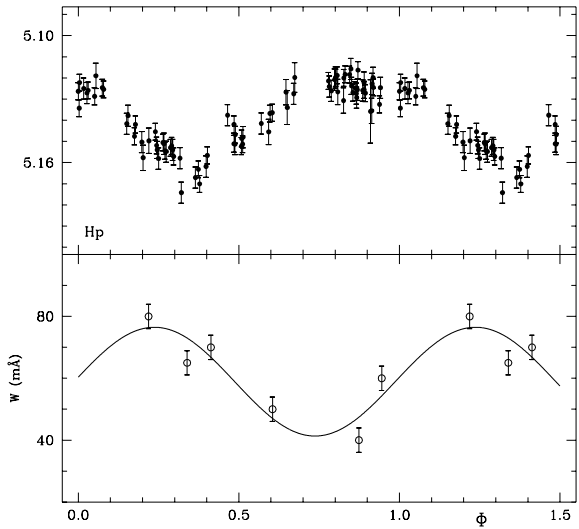,width=8cm}
  \caption{Equivalent line width variations of HD\,223640. Errorbar is 
equal to the error in the equivalent width as given by Eq.~(\ref{errorbar}). 
Photometry is from Hipparcos (ESA, 1997).}
  \label{hd223640}
 \end{figure}

\section{Conclusion}
In this paper we have presented spectroscopic observations of the 
He{\sc i}5876 {\AA} line in 16 CP stars (see Table~\ref{listcp}). In 
the case of HD\,26571, HD\,28843, HD\,43819, HD\,171247 and HD\,223640 the 
literature period values are accurate 
enough to represent our observations quite well. No variability has 
been detected in the stars HD\,22920, HD\,24587, HD\,36589, HD\,49606, and 
HD\,209515, while the He{\sc i}5876 {\AA} line has been found to be too weak 
to be measured in the stars HD\,24155, HD\,41269 and HD\,220825. 
In the case of the remaining stars, ie. HD\,176582, 177003, and HD\,182255, 
we have refined the value of the period by using both our own and literature 
data, when available.

\begin{figure}[t]
  \psfig{file=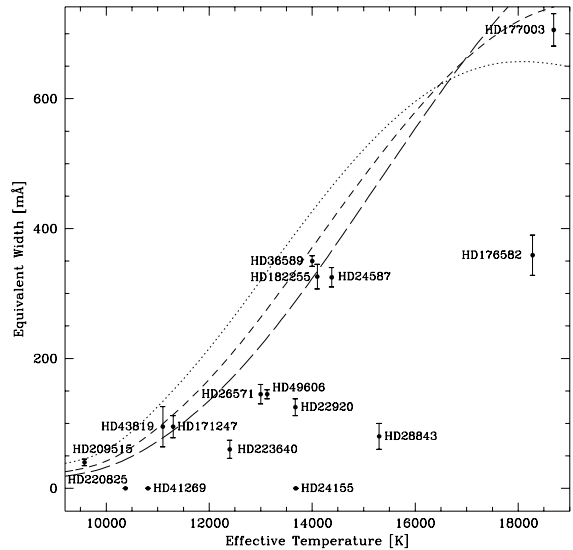,width=8cm}
  \caption{Behaviour of equivalent width versus effective temperature for our
sample of CP stars. Points represent the observations, errorbar extend by 1$\sigma$.
Curves represent the NLTE calculations by Leone \& Lanzafame (1998) for
$\log$ g: 3.5 (long dash), 4.0 (short dash) and 4.5 (dot).}
  \label{summary}
 \end{figure}

In the attempt to study the phase correlations between light and spectral
variations, we have calculated the error on phase. From these calculations
we can see that three stars, HD\,28843 (He weak), HD\,182255 (He weak) and
HD\,223640 (B9SiSrCr), show a 
clear anti-phase correlation. The equivalent width of
the He{\sc i}5876\,{\AA} line varies in phase with the photometric variations
for the stars HD\,43819 (B9 Si), HD\,171247 (B8Si) and HD\,176582 (He weak).
As to HD\,26571 the error on $\Phi$ is too large to draw any conclusion.
In the case of HD\,177003 (B3 He) nothing can be said since the Hipparcos
light curve has too large a dispersion and a low amplitude.
Hence no unique correlation exists, and this fact is 
independent of the spectral types of both groups of stars, which are 
all in the B5-B9 range. 
This result confirms the one obtained by Catalano \& Leone (1996). In an
attempt to clarify the nature of the correlation between light and helium 
lines variations, these authors compared the emerging fluxes of two
atmosphere models with the same effective temperature (T$_{eff}$\,=\,15000 K)
and gravity (log g\,=\,4.0) but with different helium abundance. The models
were computed by means of the ATLAS9 code (Kurucz 1993) and are
characterized by solar and zero helium abundance. By comparing these fluxes,
they found no observable magnitude differences and concluded that the photometric 
variations presented by helium weak stars cannot be entirely ascribed to the 
non homogeneous distribution of helium on the stellar surface.\\
Figure~\ref{summary} shows the measured average value of the equivalent 
width of the He{\sc i}5876\,{\AA} line together with the theoretical 
behaviour computed in the NLTE approximation by Leone \& Lanzafame (1998)
for solar composition stars. We conclude the helium abundance is not 
peculiar for the stars HD\,36589, HD\,43819, HD\,171247, HD\,177003, 
HD\,182255 and HD\,209515, while helium is underabundant in the 
remaining stars. It is worthy note, that the equivalent width of the
He{\sc i}5876\,{\AA} line for HD\,182255 is close to the value of a solar
composition main sequence star, even though this star is classified as an
helium weak star.  

\acknowledgements{This research has made use of the Simbad database, 
operated at CDS, Strasbourg, France.\\
We would like to thank the referee, Dr. P. Martinez, for his useful suggestions
that improved this paper.}

\end{document}